\documentclass[aps,prd,superscriptaddress,twocolumn,nofootinbib,preprintnumbers]{revtex4-2}

\usepackage{xcolor}
\usepackage[colorlinks=true,linkcolor=blue,citecolor=blue,urlcolor=blue]{hyperref}
\usepackage{amsfonts,amssymb,amsmath}
\usepackage{graphicx}

\newcommand{\result}[1]{#1}

\usepackage[normalem]{ulem}

\begin{document}
%-------------------------------------------------

\title{
When to Sweat the Small Stuff: identifying the most informative\\ events from ground-based gravitational-wave detectors
{\normalsize
}
}

\author{Reed Essick}
\email{essick@cita.utoronto.ca}
\affiliation{Canadian Institute for Theoretical Astrophysics, 60 St. George St, Toronto, Ontario M5S 3H8}
\affiliation{Department of Physics, University of Toronto, 60 St. George Street, Toronto, ON M5S 1A7}
\affiliation{David A. Dunlap Department of Astronomy, University of Toronto, 50 St. George Street, Toronto, ON M5S 3H4}

\author{Daniel E. Holz}
%\email{holz@uchicago.edu}
\affiliation{Department of Physics, Department of Astronomy \& Astrophysics, Enrico Fermi Institute, and Kavli Institute for Cosmological Physics, The University of Chicago, 5640 South Ellis Avenue, Chicago, Illinois 60637, USA}

\begin{abstract}
    We explore scaling relations for the information carried by individual events, and how that information accumulates in catalogs like those from ground-based gravitational-wave detectors.
    For a variety of situations, the larger number of quiet/distant signals dominates the overall information over the fewer loud/close sources, independent of how many model parameters are considered.
    We consider implications for a range of astrophysical scenarios, including calibration uncertainty and standard siren cosmology.
    However, the large number of additional events obtained by lowering the detection threshold can rapidly increase costs.
    We introduce a simple analysis that balances the costs of analyzing increasingly large numbers of low information events against retaining a higher threshold and running a survey for longer.
    With the caveat that precise cost estimates are difficult to determine, current economics favor analyzing low signal-to-noise ratio events.
    However, the higher detection rates expected for next-generation detectors may argue for a higher signal-to-noise ratio threshold for optimal scientific return.
\end{abstract}

\maketitle

%-------------------------------------------------

\section{Introduction}
\label{sec:introduction}

Common questions that arise when combining information from many separate observations are, ``which observations contain the most information?'' and ``to what extent does this depend on the type of measurement being performed?''
Both considerations are important for forecasts in general, but we focus on catalogs of compact binary coalescences (CBCs) observed with current (e.g., LIGO~\cite{LIGO}, Virgo~\cite{Virgo}, and KAGRA~\cite{KAGRA}) and proposed (e.g., Cosmic Explorer~\cite{CE} and Einstein Telescope~\cite{ET}) ground-based gravitational-wave (GW) detectors.
With large catalogs and finite computational and/or follow-up resources, knowing how information is encoded within the observations can help limit the cost required to reach a target precision.

This manuscript reviews how information from multiple events can be combined and establishes basic rules-of-thumb to guide cost-benefit analyses.
In particular, it corrects previous claims in the literature about the impact of the number of parameters constrained simultaneously~\cite{Haster:2020sdh}, and provides improved guidelines for which events contribute to a given measurement (Sec.~\ref{sec:stacking}).
It then provides a few examples in Sec.~\ref{sec:complicated scalings}, concluding with a discussion of cost optimization with current and proposed detectors in Sec.~\ref{sec:discussion}.
Large sets of samples used in all figures (both astrophysical and detected distributions) are available in~\citet{essick_2024_12546536}.

%-------------------------------------------------

\section{Combining Observations}
\label{sec:stacking}

Our analysis of CBCs' information content is based on the Fisher information matrix~\cite{Fisher:1922},
\begin{equation}
    \Gamma_{\alpha \beta} \equiv \int dx\, p(x|\theta) \frac{\partial \ln p(x|\theta)}{\partial \theta_\alpha} \frac{\partial \ln p(x|\theta)}{\partial \theta_\beta},
\end{equation}
which provides an estimate of the constraining power of a single observation (with data $x$) on source parameters ($\theta$) given a likelihood $p(x|\theta)$.
Specifically, we investigate how $\Gamma$ scales with the signal-to-noise ratio ($\rho$) of individual events and the distribution of $\rho$ within GW catalogs.

For high-$\rho$ signals, one can approximate the likelihood as Gaussian with precision (inverse covariance) matrix $\Gamma$~\cite{Cutler:1994ys, Vallisneri:2007ev}.
As such, a convenient way to quantify the information carried by an event is the determinant of $\Gamma$, which serves as a proxy for the (inverse of the) hypervolume in $\theta$-space that is consistent with observations.
Larger $\det|\Gamma|$ imply a tighter constraint and more information about $\theta$.

The joint likelihood of a catalog is the product of the individual events' likelihoods (assuming independent noise), each of which can be approximated as a Gaussian with Fisher matrix\footnote{Importantly, this model assumes the parameters are shared by all events, like the graviton's mass or (approximately) the neutron star radius as opposed to, e.g., the mass of each component in a binary. This caveat also applies to the analysis within, e.g.,~\citet{Haster:2020sdh}. Expressions for the Fisher matrix for more general hierarchical inferences also exist~\cite{Gair:2022fsj}, but they cannot be used within Eq.~\ref{eq:5}.} $\Gamma^{(e)}_{\alpha\beta}$.
That is,
\begin{align}
    \ln p(\{x_e\}|\theta)
        & = \sum\limits_e^{N_e} \ln p(x_e|\theta) \nonumber \\
        & \sim -\frac{1}{2}\sum\limits_e^{N_e} \theta_\alpha \Gamma^{(e)}_{\alpha\beta} \theta_\beta \nonumber \\
        & \sim -\frac{1}{2} \theta_\alpha \left(\sum\limits_e^{N_e} \Gamma^{(e)}_{\alpha\beta}\right) \theta_\beta \label{eq:5},
\end{align}
and the total precision matrix is the sum of the Fisher matrixes from each event: $\Gamma^{(\mathrm{tot})}_{\alpha\beta} = \sum_e \Gamma^{(e)}_{\alpha\beta}$.
If one quantifies the total information as the hypervolume to which the parameters are constrained by the joint analysis, one should consider the determinant of the total precision matrix $\Gamma^{(\mathrm{tot})}$,
\begin{align}
    \mathcal{I}_{\det\Sigma} & = \det\left|\sum\limits_e^{N_e} \Gamma^{(e)}_{\alpha\beta}\right| \nonumber \\
                & \approx \det\left| N_e \int\limits_{\rho_\mathrm{thr}}^{\rho_\mathrm{max}} d\rho\, p(\rho) \Gamma_{\alpha\beta}\right| \label{eq:joint info},
\end{align}
where $p(\rho)$ is the distribution of $\rho$ expected for a catalog of detected events.
For a Euclidean universe (or nearby sources)~\cite{Schutz:2011tw, Chen:2014yla},
\begin{equation} \label{eq:snr distrib}
    p(\rho) = \frac{\Theta(\rho_\mathrm{thr} \leq \rho)}{\rho^4}\left(\frac{3\rho_\mathrm{max}^3\rho_\mathrm{thr}^3}{\rho_\mathrm{max}^3 - \rho_\mathrm{thr}^3}\right),
\end{equation}
where $\rho_\mathrm{thr}$ is the detection threshold, $\Theta$ is the indicator function, and one typically takes $\rho_\mathrm{max} \rightarrow \infty$.
This distribution also implicitly assumes that we have a signal-dominated catalog (i.e., noise events are rare), which will not be true if $\rho_\mathrm{thr}$ is lowered arbitrarily.
For CBC parameters that primarily affect the GW phase and stationary Gaussian noise, $\Gamma_{\alpha\beta} \sim \rho^2$, which yields
\begin{equation}\label{eq:correct scaling}
    \mathcal{I}_{\det\Sigma} \propto \left[ N_e \left(\frac{1}{\rho_\mathrm{thr}} - \frac{1}{\rho_\mathrm{max}}\right) \left(\frac{3\rho_\mathrm{max}^3\rho_\mathrm{thr}^3}{\rho_\mathrm{max}^3 - \rho_\mathrm{thr}^3}\right)\right]^D.
\end{equation}
It is interesting to note that $\mathcal{I}_{\det\Sigma}$ is not affected if one asks unrelated questions separately or simultaneously (it scales in the same way as the product of separate questions). See Appendix~\ref{sec:haster} for more discussion.

Eq.~\ref{eq:joint info} can also incorporate (certain types of) prior information trivially.
Defining the information as the amount of hypervolume that is viable \textit{a posteriori} and including a Gaussian prior with covariance $P_{\alpha\beta}$, the joint posterior becomes
\begin{equation}
    \ln p(\theta|\{x_e\}) \sim -\frac{1}{2} \theta_\alpha \left( P^{-1}_{\alpha\beta} + \sum\limits_e^{N_e} \Gamma^{(e)}_{\alpha\beta} \right) \theta_\beta,
\end{equation}
in which case the relevant information metric is
\begin{equation}\label{eq:info a posteriori}
    \mathcal{I} = \det\left| P^{-1}_{\alpha\beta} + \sum\limits_e^{N_e} \Gamma^{(e)}_{\alpha\beta} \right|.
\end{equation}
If the Fisher matrix elements are small (i.e., weak constraints from individual events), the (inverse) prior can dominate the sum.
However, a sufficient number of signals will eventually overwhelm the prior, and the overall constraint will be dominated by the quietest signals in the catalog in the long run.

\begin{figure*}
    \begin{center}
        \includegraphics[width=1.0\textwidth]{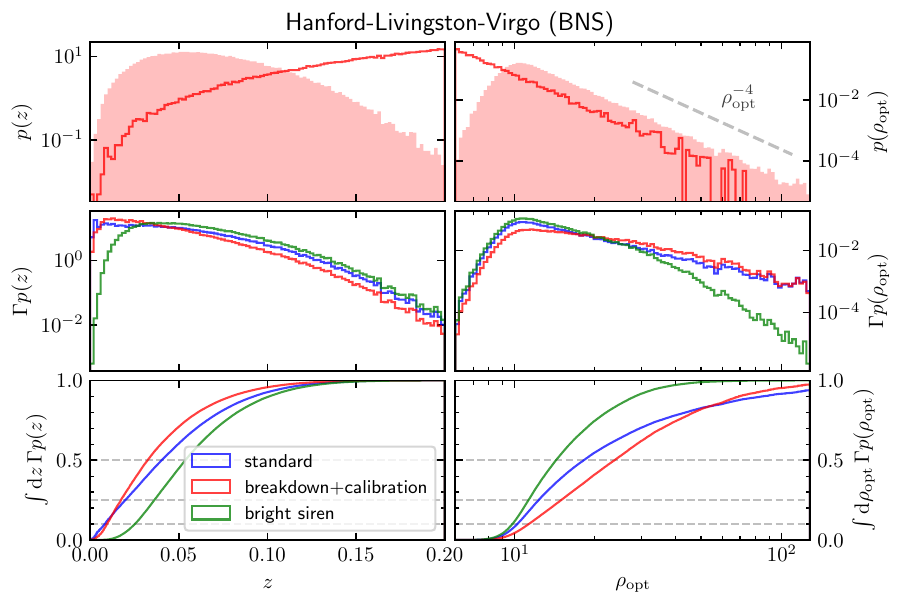}
    \end{center}
    \vspace{-0.5cm}
    \caption{
        Distributions of binary neutron star (BNS) signals and their contributions to $\mathcal{I}_{\det\Sigma}$ over the redshift to the source (\emph{left}, $z$) and the source's optimal signal-to-noise ratio (\emph{right}, $\rho_\mathrm{opt}$) with a network of advanced LIGO and Virgo detectors at design sensitivity.
        (\emph{top to bottom}) Distributions of astrophysical (\emph{solid}) and detected (\emph{shaded}) signals (with observed signal-to-noise ratio $\geq 10$~\cite{Essick:2023toz}),
        the detected-signal distribution weighted by individual events' Fisher matrixes,
        and the (normalized) integral for $\mathcal{I}_{\det\Sigma}$ showing how much information is accumulated as the upper limit of the integral is increased.
        The bottom panels show the relative importance of different events assuming
            $\Gamma \sim \rho^2$ (\emph{blue}, standard scaling),
            $\Gamma \sim \rho^4(1 + (\rho/20)^2)^{-1} (1 + (\rho/100)^2)^{-1}$ (\emph{red}, breakdown of the Fisher approximation at $\rho\lesssim 20$~\cite{Lackey:2014fwa} and pessimistic calibration uncertainty above $\rho \gtrsim 100$~\cite{Essick:2022vzl}), and
            $\Gamma \sim [((200\,\mathrm{km/s})/cz)^2 + (10\%(10/\rho))^2]^{-1}$ (\emph{green}, bright sirens)~\cite{Chen:2017rfc}.
        Horizontal lines highlight the points at which 10\%, 25\%, and 50\% of the total information is accumulated.
    }
    \label{fig:information density}
\end{figure*}

%-------------------------------------------------

\section{More Complicated $\Gamma$ vs. $\rho$ Scalings}
\label{sec:complicated scalings}

The quietest events carry a lot of information in aggregate when $\Gamma \sim \rho^2$.
However, in general, $\Gamma$ may have a more complicated relationship with $\rho$, in which case the total constraint may no longer be dominated by the quietest events in the catalog.
The following sections provide a few concrete examples, which are summarized in Fig.~\ref{fig:information density} for advanced LIGO and Virgo.
Appendix~\ref{sec:other populations} discusses how this behavior can differ between current and proposed GW detectors.

%------------------------

\subsection{Breakdown of the Fisher Approximation}
\label{sec:breakdown of fisher}

The assumption that the single-event precision matrix scales as $\Gamma_{\alpha\beta} \sim \rho^2$ only holds in the high-$\rho$ limit.
In truth, the Fisher information matrix only provides a lower-limit on the covariance (upper-limit on the precision matrix), and the actual covariance can be much larger~\cite{Vallisneri:2007ev}.
This happens for small $\rho$, which means that the events which contribute meaningfully to the sum in Eq.~\ref{eq:joint info} may effectively need to pass a higher threshold than detection.

For example, one possible scaling is
\begin{equation}\label{eq:this one}
    \Gamma \sim \rho^2 \left(\frac{\rho^2}{\rho_o^2 + \rho^2}\right),
\end{equation}
for which $\Gamma \sim \rho^4$ at small $\rho$ and $\Gamma \sim \rho^2$ for large $\rho$.
Here, $\rho_o$ sets the scale below which the Fisher approximation begins to break down, which will almost certainly be problem-dependent.
Additionally, the breakdown in $\Gamma \sim \rho^2$ may not exactly follow Eq.~\ref{eq:this one}, but Eq.~\ref{eq:this one} captures the general behavior and makes the associated integrals analytically tractable.
Assuming $\rho_\mathrm{max} \rightarrow \infty$,
\begin{equation} \label{eq:breakdown}
    \int dN \Gamma \sim N_e \frac{3 \rho_\mathrm{thr}^3}{\rho_o}\left(\frac{\pi}{2} - \tan^{-1}\left(\frac{\rho_\mathrm{thr}}{\rho_o}\right)\right).
\end{equation}
Since $N_e \sim \rho_\mathrm{thr}^{-3}$ for nearby sources, Eq.~\ref{eq:correct scaling} diverges as $\rho_\mathrm{thr}^{-1}$ as $\rho_\mathrm{thr} \rightarrow 0$.
However, Eq.~\ref{eq:breakdown} remains finite in the same limit.
Indeed, $\rho_o$ determines which events carry the most information; at least half the total information available is carried by the events with $\rho \geq \rho_o$ (Fig.~\ref{fig:information density}).

Additionally, a breakdown in the Fisher approximation is a likely explanation for the behavior reported in~\citet{Lackey:2014fwa} (neutron star equation of state constraints are dominated by the loudest events in their catalog).
Motivated by~\citet{Lackey:2014fwa}, Fig.~\ref{fig:information density} assumes $\Gamma \sim \rho^2$ breaks down below $\rho_o \sim 20$.

%------------------------

\subsection{Calibration Uncertainty}
\label{sec:calibration uncertainty}

Alternatively, $\Gamma \sim \rho^2$ may break down at high $\rho$ in the presence of calibration or waveform uncertainty.
\citet{Essick:2022vzl} shows that $\Gamma$ can approach a constant at large $\rho$ (e.g., the fractional uncertainty from calibration), which can be modelled as
\begin{equation}
    \Gamma \sim \rho^2 \left(\frac{\rho_o^2}{\rho_o^2 + \rho^2}\right),
\end{equation}
where $\rho_o$ defines the scale at which calibration uncertainties begin to dominate, in which case\footnote{This assumes calibration errors are independent for each event.}
\begin{equation}
    \int dN \Gamma \sim N_e \frac{3\rho_\mathrm{thr}^3}{\rho_o}\left( \frac{\rho_o}{\rho_\mathrm{thr}} + \tan^{-1}\left(\frac{\rho_\mathrm{thr}}{\rho_o}\right) - \frac{\pi}{2} \right).
\end{equation}
The quietest signals in the catalog dominate the overall information even more than for the standard scaling, although this approaches the standard scaling if $\rho_\mathrm{thr} \ll \rho_o$ (i.e., if almost all of the catalog is too quiet to be affected by calibration errors).
See Fig.~\ref{fig:information density}, which assumes calibration limits the precision of signals at $\rho_o \sim 100$ (much lower than is likely to be the case for realistic calibration uncertainties~\cite{Essick:2022vzl}).

%------------------------

\subsection{Standard Sirens}
\label{sec:standard sirens}

Finally, single-event constraints may always be a complicated function of $\rho$, and the approximation $\Gamma \sim \rho^2$ may simply never hold.
Standard sirens, where one measures the Hubble parameter ($H_0$) at the ratio of a redshift ($z$) from a (candidate) host galaxy and a luminosity distance ($D_L$) from the GW strain, are one such example.

In the simplest approximation,
\begin{equation}
    H_0 = \frac{cz}{D_L},
\end{equation}
where $c$ is the speed of light.
Standard error propagation suggests that
\begin{equation}
    \Gamma^{-1}_{H_0} \sim \sigma_{H_0}^2 \sim H_0^2 \left(\frac{\sigma_z^2}{z^2} + \frac{\sigma^2_{D_L}}{D_L^2}\right),
\end{equation}
where $z$ is estimated from a set of candidate host galaxies and $D_L$ is estimated from the GW signal.

A straightforward Fisher analysis for the GW data (neglecting the degeneracy with inclination) suggests that
\begin{equation}
    \sigma_{D_L} \propto \frac{D_L}{\rho}.
\end{equation}
Including the degeneracy with inclination, one might expect $\sigma_{D_L} \propto D_L$.
Breaking this degeneracy depends on measuring the GW polarization and/or higher-order modes, which has been done but often requires relatively large $\rho$~\cite{GW190412, GW190814}.

\begin{figure*}
    \includegraphics[width=1.0\textwidth]{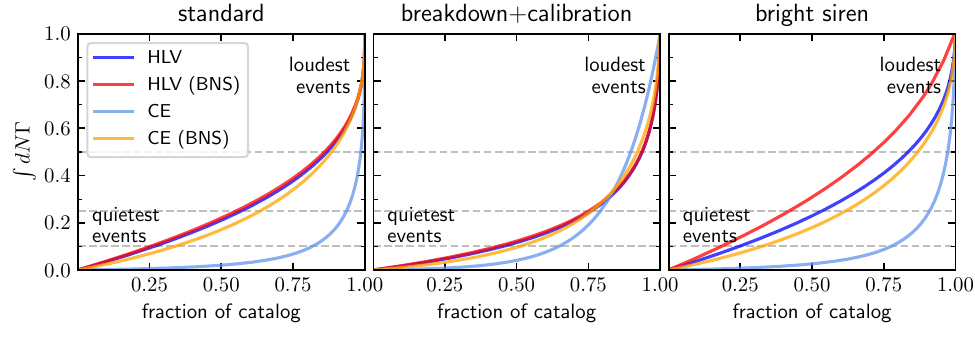}
    \caption{
        Integrated information vs. the fraction of the catalog considered (including quieter events first) when $\Gamma$ scales with $\rho$ in the ways shown in Fig.~\ref{fig:information density}.
        We show the results for low-mass signals observed with
            advanced LIGOs and Virgo (\emph{red}) or
            CE (\emph{orange}) as well as the full range of CBCs observed with advanced detectors (\emph{dark blue}) or CE (\emph{light blue}).
        Horizontal lines highlight when 10\%, 25\%, and 50\% of the total information has been accumulated.
    }
    \label{fig:info by frac}
\end{figure*}

The uncertainty in $z$ from a galaxy catalog is the sum of two components: the (average) measurement uncertainty for the redshift of individual galaxies (which could include their peculiar velocity) and a term that captures the variance of redshifts within the catalog ($\{\mu_g\}$):
\begin{equation} \label{eq:simple sigma_z^2}
    \sigma_z^2 \sim \frac{1}{N_g}\sum\limits_g^{N_g} \sigma_g^2 + \frac{1}{N_g}\sum\limits_g^{N_g} \left(\mu_g - \frac{1}{N_g}\sum\limits_g^{N_g} \mu_g\right)^2.
\end{equation}
Assuming most galaxies have similar measurement uncertainty ($\sigma_g$), only the second term depends on $\rho$ (through the set of galaxies consistent with the GW localization).
Here, we focus on the case of a bright siren, where a single host galaxy is identified ($N_g=1$ and $\sigma_z^2 = \sigma_g^2$).

As discussed in~\citet{Chen:2017rfc}, the precision on $H_0$ for bright sirens is then limited by the distance uncertainty for quiet/distant events while the precision of the loudest/nearest sources is limited by the uncertainty in the redshift measurement for individual galaxies, which does not vanish as $z \rightarrow 0$.\footnote{The second term in Eq.~\ref{eq:simple sigma_z^2} may grow rapidly at low $\rho$ (large $N_g$), and therefore the precision of (dark) standard sirens may always be limited by $\sigma_z^2$. See Appendix~\ref{sec:redshift uncertainty}.}
Therefore, the most informative events in the catalog come from intermediate $\rho$.
Fig.~\ref{fig:information density} assumes $\sigma_g \sim (200\,\mathrm{km/s})/c$ and $\sigma_{D_L}/D_L \sim 10\%$ at $\rho\sim 10$.

%-------------------------------------------------

\section{Discussion}
\label{sec:discussion}

It is useful to rewrite Eq.~\ref{eq:correct scaling} in terms of only $N_e$ so that $\mathcal{I}_{\det\Sigma} \sim N_e^{1/3}$ when $\Gamma \sim \rho^2$, which implies $\sigma \sim N_e^{-1/6}$.
As such, lowering $\rho_\mathrm{thr}$ enough to decrease $\sigma$ by a factor of $\sim 2$ will increase the catalog size (and associated computational cost) by a factor of $\sim 2^6 = 64$.
Fig.~\ref{fig:info by frac} plots the cumulative information vs. the fraction of the catalog for several detector networks and scalings of $\Gamma$ vs. $\rho$.
It is clear that a large fraction of the overall information can come from a relatively small fraction of the catalog.

In contrast, if one fixes $\rho_\mathrm{thr}$ but runs the experiment longer, a factor of $\sim 4$ larger catalog size will reduce $\sigma$ by a factor of $\sim 2$.
This begs the question of whether it could be more cost-effective to run an experiment for longer with a high threshold instead of lowering $\rho_\mathrm{thr}$.

Consider the following cost model:
\begin{equation}
    C = C_F + C_O \tau + C_A(N_e),
\end{equation}
where $C_F$, $C_O$, and $C_A(N_e)$ are the fixed cost of constructing an experiment, the marginal cost of operating the experiment (with duration $\tau$), and the cost of analyzing $N_e$ events, respectively.
For current detectors, the number of events scales approximately as $N_e \sim R\tau (\rho_\mathrm{ref}/\rho_\mathrm{thr})^3$, where $R$ is the rate of detections with a threshold $\rho_\mathrm{ref}$.
The precision of the measurement will scale as $\mathcal{I} \sim N_e / f(\rho_\mathrm{thr}) \sim R\tau \rho_\mathrm{ref}^3 / f(\rho_\mathrm{thr}) \rho_\mathrm{thr}^3$, and, in the simplest approximation (Eq.~\ref{eq:joint info}), $f \sim \rho_\mathrm{thr}^{-2}$.
Minimizing $C$ with respect to $\rho_\mathrm{thr}$ at fixed $\mathcal{I}$ yields
\begin{align}
    \left. \frac{dC}{d\rho_\mathrm{thr}} \right|_\mathcal{I}
        & = C_O \frac{\mathcal{I}}{R\rho_\mathrm{ref}^3} \frac{d}{d\rho_\mathrm{thr}}\left(\rho_\mathrm{thr}^3 f\right) + \frac{dC_A}{dN_e} \mathcal{I} \frac{df}{d\rho_\mathrm{thr}} \nonumber \\
    0   & = \frac{C_O}{R} \left(\frac{\rho_\mathrm{thr}}{\rho_\mathrm{ref}}\right)^3 - 2 \frac{dC_A}{dN_e},
\end{align}
which defines an implicit expression for the optimal $\rho_\mathrm{thr}$ in terms of $\mathcal{I}$ and marginal costs.
Once this is set, the required duration is given by $\tau \sim \mathcal{I} \rho_\mathrm{thr}^3 f / \rho_\mathrm{ref}^3 R$.

Interestingly, in the simplest model where the marginal analysis cost ($dC_A/dN_e$) is constant, the optimal $\rho_\mathrm{thr}$ does not depend on the precision $\mathcal{I}$,\footnote{Although the general optimization can depend on the amount of information per event, and the amount of information per event (at fixed $\rho$) may change between, e.g., current and next-generation detectors, it is often similar~\cite{Essick:2022vzl}.} and
\begin{equation}\label{eq:optimal threshold}
    \rho_\mathrm{thr} = \rho_\mathrm{ref} \left( \frac{2 R dC_A/dN_e}{C_O} \right)^{1/3}.
\end{equation}
This balances the cost of analyzing additional events (per unit time) and the cost of operating the instrument (per unit time).

Estimating the costs in Eq.~\ref{eq:optimal threshold} precisely is non-trivial, but even rough numbers may provide guidance for current and/or planned detectors.
In this spirit, we provide such estimates below.
However, we stress that these examples are merely illustrative, and more detailed calculations should be conducted before any policy is set by such considerations.

The current operating cost of advanced LIGO is $\sim 10^7\,\mathrm{USD}/\mathrm{yr}$~\cite{aLIGO-budget} with an expected rate of detections of $R \sim 1/\mathrm{day}$ at design sensitivity~\cite{ObservingScenarios}.
Estimating the marginal cost of analyzing an event as approximately $10\,\mathrm{hrs}$ of an analyst's time ($dC_A/dN_e \sim 300\,\mathrm{USD}/\mathrm{event}$),\footnote{An hourly wage of $30\,\mathrm{USD}$ corresponds to $6\times10^{4}\,\mathrm{USD}/\mathrm{yr}$, which is likely a low estimate of labor costs including both salary and overhead.} Eq.~\ref{eq:optimal threshold} becomes $\rho_\mathrm{thr}/\rho_\mathrm{ref} \approx 0.28$.
As such, the relatively low rate of detections drives the most efficient threshold lower than the nominal value (assuming the Fisher approximation holds at low $\rho$).

However, the rate of detections may dramatically increase with next-generation detectors, by {\result{a factor of $\gtrsim 100$}}.\footnote{Although the detection horizon may increase by an order of magnitude (naively suggesting an increase of $\gtrsim 10^3$ in detection rate), this extends beyond the peak of star formation.}
Assuming $C_O$ and $dC_A/dN_e$ are roughly the same for future detectors, the optimal $\rho_\mathrm{thr}$ will be a {\result{factor of 1.4 higher than $\rho_\mathrm{ref}$}}.\footnote{If operating costs increase proportional to the detector's arms' lengths (i.e., by an order of magnitude), then {\result{$\rho_\mathrm{thr}/\rho_\mathrm{ref} \sim 0.66$}}.}
Again, this also assumes that the Fisher approximation holds for low-$\rho$ events (it almost certainly will not) and that the catalog is flux-limited.
Alternatively, if $dC_A/dN_e$ improves even moderately (by as little as {\result{a factor of $\sim 3$}}), then the optimal $\rho_\mathrm{thr}$ will still be $\lesssim \rho_\mathrm{ref}$.
This modest reduction in analysis costs may be readily achievable before next-generation detectors come online.
Additionally, the rate of binary neutron star (BNS) detections may continue to be much smaller than the overall detection rate.
This would almost guarantee that the optimal $\rho_\mathrm{thr}$ for questions involving BNS will remain $\lesssim \rho_\mathrm{ref}$ even with higher detection rates in next-generation catalogs.

All this being said, Fig.~\ref{fig:info by frac} shows that, while the simple scaling motivated by Eq.~\ref{eq:correct scaling} may approximately hold for low-mass CBCs observed with next-generation detectors, it is unlikely to hold for high-mass CBCs.
Such surveys will not be flux-limited, and the distribution $p(\rho)$ will be fundamentally different than Eq.~\ref{eq:snr distrib}.
Indeed, one will likely lose even less information by considering only the loudest high-mass events observed with CE.
See Appendix~\ref{sec:other populations} for more discussion.

Furthermore, $dC_A/dN_e$ is almost certainly not constant.
For example, contamination from noise events will grow rapidly with decreasing $\rho_\mathrm{thr}$, and the overall number of events that would require follow-up would be much larger than the number of signals~\cite{Lynch:2018}.
The confusion between noise and signal is also bound to reduce the information per signal at low $\rho$.
This effectively sets the detection threshold within current catalogs (e.g., GWTC-3~\cite{GWTC-3}).

Putting such concerns aside, one might additionally optimize over facilities by first optimizing over ($\tau, \rho_\mathrm{thr}$) separately, and then optimizing over facilities while taking into account their $C_F$.
This is often done only in the context of the science accessible with different detector designs, neglecting their relative costs (see, e.g., Refs.~\cite{Finn:1992xs, Srivastava:2022slt}).
However, a prerequisite for such an analysis would be an enumeration of science targets, and it is difficult to concisely capture the full range of discovery potential enabled by, e.g., next-generation GW detectors.
At the very least, such an exercise is beyond the scope of this paper.

Our results suggest that a more detailed cost-benefit analysis may help determine the optimal balance of survey duration and detection threshold for next-generation detectors.
The simplest scaling and cost models suggest that it makes sense to go deep within current catalogs (although our analysis is intentionally only a rough estimate).
With reasonable improvements in analysis costs, this may remain true of future catalogs even with much higher detection rates.

%-------------------------------------------------

\acknowledgments

We are grateful for computational resources provided by the LIGO Laboratory and supported by National Science Foundation Grants PHY-0757058 and PHY-0823459.

R.E. is supported by the Natural Sciences \& Engineering Research Council of Canada (NSERC) through a Discovery Grant (RGPIN-2023-03346).
D.E.H. is supported by NSF grants AST-2006645 and PHY2110507, as well as by the Kavli Institute for Cosmological Physics through an endowment from the Kavli Foundation and its founder Fred Kavli.

This study would not have been possible without the following software: \texttt{numpy}~\citep{numpy}, \texttt{scipy}~\citep{scipy}, and \texttt{matplotlib}~\citep{matplotlib}.

The large sample of detected events for both HLV at design sensitivity and CE, including the impact of the frequency-dependent interferometric response~\cite{Essick:2017wyl}, are publicly available in~\citet{essick_2024_12546536}.
Samples were generated with \texttt{monte-carlo-vt}~\cite{monte-carlo-vt}, \texttt{gw-distributions}~\cite{gw-distributions}, and \texttt{gw-detectors}~\cite{gw-detectors}.

%-------------------------------------------------

\appendix

%-------------------------------------------------

\section{A Comment on~\citet{Haster:2020sdh}}
\label{sec:haster}

Previously,~\citet{Haster:2020sdh} argued that the dimensionality of the question(s) being asked plays a role in which events carry the most information.
That is, they claim that the set of events which are most informative depends on the number of parameters being constrained: quiet/distant events dominate one-dimensional questions but higher-dimensional questions depend most strongly on loud/nearby events.
However, if one considers two properties that are unrelated (i.e., uncorrelated at the single-event level), the amount of information, and which events carry that information, should not depend on whether these properties are estimated separately or simultaneously.

\citet{Haster:2020sdh} correctly identify that each element of $\Gamma$ often scales with $\rho^2$ for a single event.
As such, $\det|\Gamma|$ should have an overall scaling of $\rho^{2D}$, where $D = \mathrm{rank}\{\Gamma\}$ is the dimensionality of the problem.
They then estimate the total information by summing $\det|\Gamma|$ from separate events.
That is, they compute
\begin{align}
    \mathcal{I}_{\Sigma\det} & = \sum\limits_e^{N_e} \det\left|\Gamma^{(e)}_{\alpha\beta}\right| \nonumber \\
                & \approx N_e \int\limits_{\rho_\mathrm{thr}}^{\rho_\mathrm{max}} d\rho\, p(\rho) \det\left|\Gamma_{\alpha\beta}\right| \label{eq:joint info haster}
\end{align}
which has very different behavior than Eq.~\ref{eq:joint info}.
If $\det|\Gamma_{\alpha\beta}|$ scales as $\rho^{2D}$, then Eq.~\ref{eq:joint info haster} becomes
\begin{equation}
    \mathcal{I}_{\Sigma\det} \propto N_e \left(\frac{\rho_\mathrm{max}^{2D-3} - \rho_\mathrm{thr}^{2D-3}}{2D-3}\right) \left(\frac{3\rho_\mathrm{max}^3\rho_\mathrm{thr}^3}{\rho_\mathrm{max}^3 - \rho_\mathrm{thr}^3}\right)
\end{equation}
which diverges as $\rho_\mathrm{max} \rightarrow \infty$ if $D \geq 3/2$.
This logic suggests that the single loudest event carries almost all the information for high-dimensional questions ($D \gg 1$).

$\mathcal{I}_{\Sigma\det}$ also suggests that the constraint on each parameter scales worse with the size of the catalog if multiple questions are asked simultaneously ($\mathcal{I}^{1/D}_{\Sigma\det} \sim N^{1/D}$), even if they are unrelated.
That is, even if $\Gamma_{\alpha\beta}^{(e)}$ is always block-diagonal,~\citet{Haster:2020sdh} incorrectly predict that asking multiple questions simultaneously affects our ability to answer them.

Fundamentally, then, the issue with the analysis in~\citet{Haster:2020sdh} boils down to the fact that
\begin{equation}
    \sum\limits_e^{N_e} \det\left|\Gamma^{(e)}_{\alpha\beta}\right| \neq \det\left|\sum\limits_e^{N_e} \Gamma^{(e)}_{\alpha\beta}\right|
\end{equation}
Many of their conclusions still hold, though.
Their prediction that the cosmologically-distant sources at $z \sim 1$ will dominate the total information from BNS mergers observed with next-generation GW detectors is correct (most of the observed sources will be at $z \sim 1$; see Appendix~\ref{sec:other populations}).
Similarly, they correctly point out that tidal information at high frequencies will be redshifted into the sensitive band of GW detectors for cosmologically distant BNS (see also Refs.~\cite{Chatterjee:2021xrm, Xie:2022brn}), and this makes $\Gamma$ a more complicated function of $\rho$ (other examples are given in Sec.~\ref{sec:complicated scalings}).
In this way,~\citet{Haster:2020sdh} correctly point out that the set of events which contain the most information can depend on the behavior of $\Gamma(\rho)$, even though they incorrectly attribute this to the dimensionality of the question.
Lastly, their observation that the BNS merger rate may not scale with the comoving volume is also correct.

%------------------------

\section{Other Populations and Detector Networks}
\label{sec:other populations}

This appendix compares the behavior of different populations in different detector networks.

\begin{figure*}
    \begin{minipage}{0.49\textwidth}
        \includegraphics[width=1.0\textwidth]{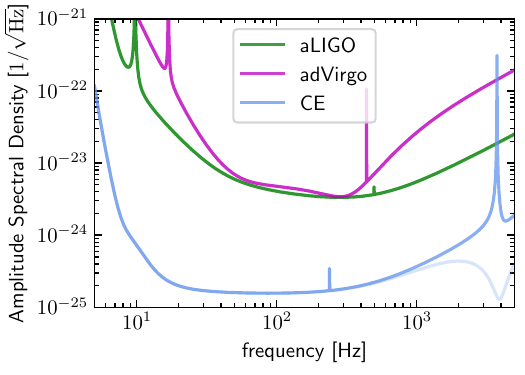}
    \end{minipage}
    \begin{minipage}{0.49\textwidth}
        \includegraphics[width=1.0\textwidth]{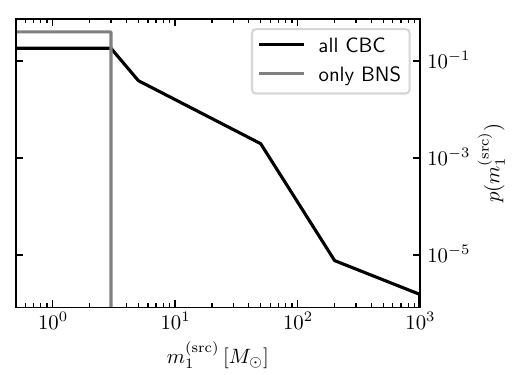}
    \end{minipage}
    \caption{
        (\emph{left}) Amplitude spectral densities for advanced LIGO and Virgo at design sensitivity and Cosmic Explorer.
        For CE, the dark line shows the response to a source that is directly overhead the detector, and light line shows the response without frequency-dependent interferometric effects~\cite{Essick:2017wyl}.
        (\emph{right}) Source-frame primary mass ($m_1^{(\mathrm{src})}$) distributions used within this study, intended to mimic the observed distribution of merging binaries (all CBC~\cite{GWTC-3-RnP}) or capture the behavior of only low-mass objects (only BNS).
    }
    \label{fig:PSDs and mass distributions}
\end{figure*}

Fig.~\ref{fig:PSDs and mass distributions} shows the one-sided amplitude spectral densities (ASDs) used for the two LIGO detectors (Hanford and Livingston), Virgo, and CE.
It also shows the two source-frame primary-mass distributions considered: one that is based roughly on the current observed population of merging binaries~\cite{GWTC-3-RnP} and one restricted to low-masses intended to model BNS systems ($1\,M_\odot \leq m_2^{(\mathrm{src})} \leq m_1^{(\mathrm{src})} \leq 3\,M_\odot$).

\begin{figure*}
    \begin{minipage}{0.49\textwidth}
        \includegraphics[width=1.0\columnwidth]{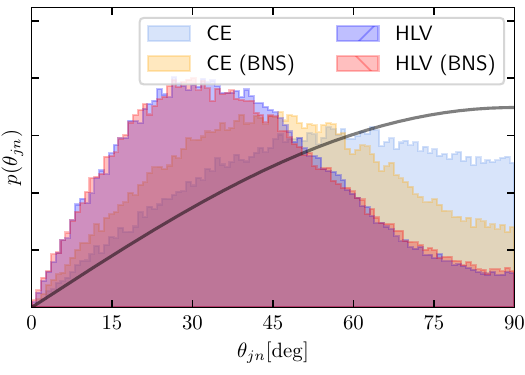}
    \end{minipage}
    \begin{minipage}{0.49\textwidth}
        \includegraphics[width=1.0\columnwidth]{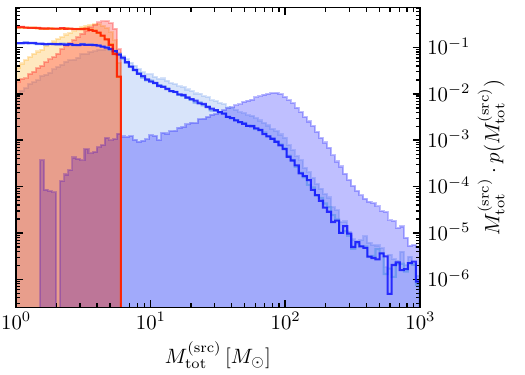}
    \end{minipage}
    \caption{
        (\emph{left}) Distribution of orbital inclinations ($\theta_{jn}$) for detected events with both populations from Fig.~\ref{fig:PSDs and mass distributions} for HLV at design sensitivity and CE.
        Because the distribution is symmetric about $\theta_{jn} = 90^\circ$, we only show $0^\circ \leq \theta_{jn} \leq 90^\circ$.
        An isotropic distribution (\emph{black line}) is also shown for reference.
        (\emph{right}) Distributions of total source-frame mass ($M_\mathrm{tot}^{(\mathrm{src})}$) for
            the astrophysical (\emph{solid lines}) and
            detected (\emph{shaded}) populations.
        While the selection in favor of high-mass systems is reduced at high masses for CE, it remains for $M_\mathrm{tot}^{(\mathrm{src})} \lesssim 5\,M_\odot$.
    }
    \label{fig:inclinations}
\end{figure*}

Fig.~\ref{fig:inclinations} shows the distribution of inclinations ($\theta_{jn}$) for detected events with both populations shown in Fig.~\ref{fig:PSDs and mass distributions} for both the HLV network and for CE.
Although the selection against edge-on binaries ($\theta_{jn} = 90^\circ$) is reduced for CE compared to HLV, it is still noticeable.
However, unlike with HLV, it is also clear that the extent of this selection depends on the binary's masses with CE.
Indeed, Fig.~\ref{fig:inclinations} also shows that there is still a strong selection towards more massive binaries within low-mass populations.
Catalogs of low-mass systems from next-generation detectors will still be flux-limited and have strong selections on, e.g., $\theta_{jn}$ and the component masses.\footnote{The conclusions in, e.g.,~\citet{Vitale:2016aso} are only true for massive binary black holes (BBHs).~\citet{Vitale:2016aso} focuses on BBHs and acknowledge that their observations may not apply for low-mass systems. See also~\citet{Vitale:2023}.}

\begin{figure*}
    \begin{center}
        \includegraphics[width=1.0\textwidth]{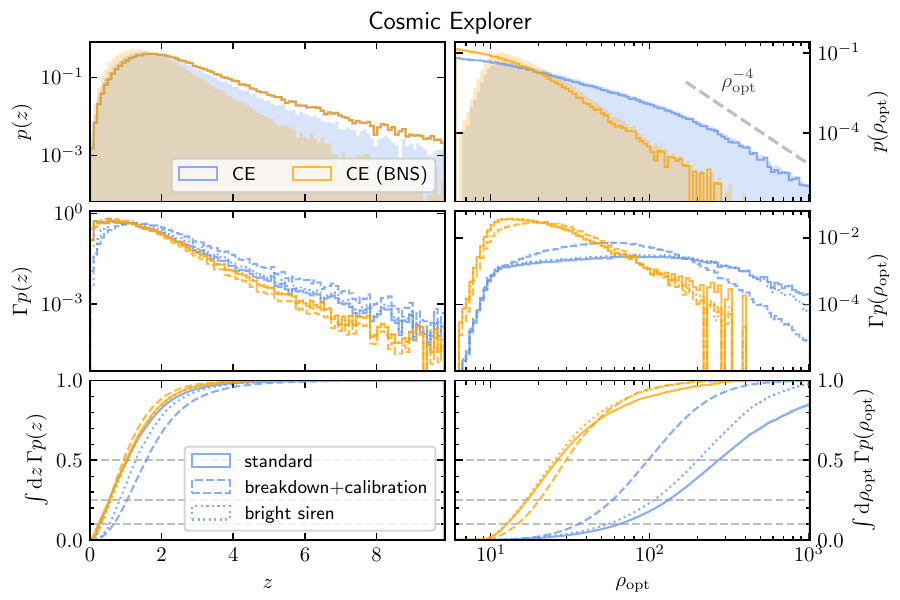}
    \end{center}
    \vspace{-0.5cm}
    \caption{
        Analogous to Fig.~\ref{fig:information density} but for sets of events detected with CE.
        The cosmological reach of next-generation detectors (to beyond the peak of star formation) means that the distribution of detected $\rho$ falls off much less steeply for high-mass systems.
        As such, relatively little information is lost by neglecting the low-$\rho$, high-mass systems.
        However, catalogs of low-mass systems remain flux-limited, and a significant fraction of the overall information available is carried by the low-$\rho$, low-mass events.
    }
    \label{fig:CE information density}
\end{figure*}

Finally, Fig.~\ref{fig:CE information density} shows which events are most important with catalogs of sources detected by CE (compare to Fig.~\ref{fig:information density}).
The main difference boils down to $p(\rho)$, which, again, is not flux-limited for massive BBH in CE.
As such, some of the scalings reported in the main text break for high-mass signals in CE.
Indeed, {\result{one may increase $\rho_\mathrm{thr}$ from $\sim 10$ up to $\sim 100$ without losing much information if $\Gamma \sim \rho^2$ for high-mass CBCs}}.
However, {\result{this would correspond to a reduction of a factor of only $\sim 10$ in the catalog size (Fig.~\ref{fig:info by frac}), rather than the expected factor of $10^3$ for a flux-limited catalog}}.

%------------------------

\section{Statistical Standard Siren Redshift Uncertainty}
\label{sec:redshift uncertainty}

The distribution of redshifts from a galaxy catalog can be approximated as
\begin{equation}
    p(z) = \sum\limits_g^{N_g} \frac{w_g}{\sqrt{2\pi\sigma_g^2}} e^{-(z-\mu_g)^2/2\sigma_g^2}
\end{equation}
assuming Gaussian errors on each galaxy's measured redshift and a weight
\begin{equation}
    w_g \sim p(\Omega_g|x_\mathrm{GW}) \omega_g
\end{equation}
for each galaxy, which is the product of the consistency of the Galaxy's position on the sky ($\Omega_g$) with the GW data ($x_\mathrm{GW}$) and an additional weight based on the galaxy's properties ($\omega_g$), which captures prior beliefs about the types of galaxies that are more likely to host binary mergers (see, e.g.,~\citet{Hanselman:2024hqy}).
With such a catalog, one can immediately compute
\begin{equation}\label{eq:sigma_z^2}
    \sigma_z^2 \equiv \mathrm{V}[z] = \sum\limits_g^{N_g} w_g \sigma_g^2 + \sum\limits_g^{N_g} w_g \left(\mu_g - \sum\limits_g^{N_g} w_g \mu_g\right)^2
\end{equation}
which is Eq.~\ref{eq:simple sigma_z^2} when $w_g = 1/N_g$.

One can estimate $\sigma_z^2$ for a representative catalog by taking the expected value of Eq.~\ref{eq:sigma_z^2} with respect to set of galaxies that are uniformly distributed in co-moving volume with a Poisson distribution for the number of galaxies in the catalog, neither of which are perfect models of reality.
That is, for simplicity, assume
\begin{equation}
    w_g = \frac{1}{N_g}
\end{equation}
and
\begin{equation}
    p(\mu_g) \propto \frac{d V_c}{dz}
\end{equation}
for $z_\mathrm{min} < z < z_\mathrm{max}$.
This yields
\begin{equation}
    \mathrm{E}[\sigma_z^2]_{\mu_g|N_g} = \frac{1}{N_g}\sum\limits_g \sigma_g^2 + \left(1 - \frac{1}{N_g} \right)\mathrm{V}[\mu_g]
\end{equation}
where
\begin{equation}
    \mathrm{V}[\mu_g] = \int\limits_{z_\mathrm{min}}^{z_\mathrm{max}} dz p(z) z^2 - \left(\int\limits_{z_\mathrm{min}}^{z_\mathrm{max}} dz p(z) z\right)^2
\end{equation}
For well-localized events, $N_g \rightarrow 1$ and the second term vanishes.
For poorly localized events, $N_g \gg 1$ and the second term may dominate.

Taking the expectation value with respect to the number of galaxies, which is Poisson-distributed
\begin{equation}
    P(N_g) = \frac{\Lambda^{N_g}}{N_g!} \left(\frac{1}{e^\Lambda - 1} \right)
\end{equation}
for $N_g = 1, 2, \ldots$, yields
\begin{align}
    \mathrm{E}[\sigma_z^2]_{\mu_g, N_g}
        & = \left<\sigma_g^2\right> + \mathrm{V}[\mu_g]\left( 1 - \sum\limits_{N_g=1}^\infty P(N_g) \frac{1}{N_g} \right) \nonumber \\
        & = \left<\sigma_g^2\right> + \mathrm{V}[\mu_g]\left( 1 - \frac{1}{e^\Lambda -1}\int\limits_0^\Lambda d\lambda \frac{e^\lambda - 1}{\lambda}\right)
\end{align}
Although the integral is difficult to express in closed form, one can still gain insight from recognizing that the second term approaches $\mathrm{V}[\mu_g] \Lambda / 4$ as $\Lambda \rightarrow 0$, and that
\begin{equation}
    \mathrm{E}[N_g] = \Lambda \propto (\Delta \Omega) \left(\frac{dN_g}{dz} \Delta z\right)
\end{equation}
could scale in several ways.
$\Delta \Omega$ should scale as $\rho^{-1}$ ($\rho^{-2}$) for 2-IFO (3+ IFO) localizations, and $\Delta z$ should scale as $\Delta z = z_\mathrm{max} - z_\mathrm{min} \sim (D_L/c)\Delta H_0 \sim \rho^{-1}$ where $\Delta H_0$ is the extent of the $H_0$ prior, although it could also scale as a more complicated function of $\rho$ if one took into account the uncertainty on $D_L$.

Similarly, if galaxies are uniformly distributed in volume, then $dN_g/dz \sim z^2 \sim \rho^{-2}$ (for nearby sources).
As such, one may expect $\mathrm{E}[N_g]$ to scale strongly with $\rho$, and the uncertainty $(1-N_g^{-1})\mathrm{V}[\mu_g]$ may shrink rapidly for loud signals.
Additionally, $\mathrm{V}[\mu_g]$ may scale as $(\Delta z)^2 \sim \rho^{-2}$ when $N_g \gg 1$, so that $\mathrm{V}[\mu_g]/z^2 \sim \mathrm{constant}$.

%-------------------------------------------------

\bibliography{biblio.bib}

%-------------------------------------------------
\end{document}